\documentclass[12pt]{iopart}
\usepackage{graphicx,xcolor}
\usepackage{dsfont}

\newcommand{\heff}{H_{\scriptscriptstyle \rm eff}}
\newcommand{\Heff}{H^I_{\scriptscriptstyle \rm eff}}
\newcommand{\hs}{\mathcal{H}_{\scriptscriptstyle \rm S}}

\newcommand{\hl}{\mathcal{H}_{\scriptscriptstyle \rm L}}

\newcommand{\hsl}{\mathcal{H}_{\scriptscriptstyle \rm SL}}

\newcommand{\hdd}{\mathcal{H}_{\scriptscriptstyle \rm DD}}

\newcommand{\nn}{\nonumber}

\newcommand{\rh}{\rho_{\scriptscriptstyle  S}}
\newcommand{\rl}{\rho_{\scriptscriptstyle \rm L}^{\scriptscriptstyle \rm eq}}

\newcommand{\stoo}{\vert_{t\rightarrow 0}}

\begin{document}
\title{ Emergence of superradiance in dissipative dipolar-coupled spin systems }

\author{Saptarshi Saha$^1$}
\eads{\mailto{ss17rs021@iiserkol.ac.in}}

\author{Yeshma Ibrahim$^2$}
\eads{\mailto{yeshma@phy.iitb.ac.in}}

\author{Rangeet Bhattacharyya$^1$}
\eads{\mailto{rangeet@iiserkol.ac.in}}

\address{$^1$Department of Physical Sciences, 
Indian Institute of Science Education and Research Kolkata,
Mohanpur - 741 246, West Bengal, India\\
$^2$Department of Physics, 
Indian Institute of Technology Bombay,
Powai - 400 076, Mumbai, India }

\begin{abstract}

In the superradiance phenomenon, a collection of non-interacting atoms exhibits collective dissipation due to
interaction with a common radiation field, resulting in a non-monotonic decay profile. This work shows that dissipative dipolar-coupled systems exhibit an identical collective dissipation aided by the nonsecular
part of the dipolar coupling.  We consider a simplified dipolar network where the dipolar interaction between
the spin-pairs is assumed to be identical. Hence the dynamics remain confined in
the block diagonal Hilbert
spaces. For a suitable choice of the initial condition, the resulting dynamics require dealing
with a smaller subspace which helps extend the analysis to a larger spin network. To include the nonsecular
dipolar relaxation, we use a fluctuation-regulated quantum master equation. We note that a successful
observation of superradiance in this system requires a weak system-bath coupling. 
Moreover, we find that for an ensemble of $N$ spins, the maximum intensity of the radiation exhibits a
nearly quadratic scaling ($N^2$), and the dipolar relaxation time follows an inverse square proportionality
($1/N^2$); these two observations help characterize the emergence of superradiance. Our results agree well
with the standard results of pure spin superradiance observed experimentally in various systems. 

\end{abstract}
\maketitle

\section{Introduction}\label{intro}

The collective behavior of interacting many-particle systems can deviate significantly from the individual particles \cite{doi:10.1126/science.177.4047.393}. The observation of emergent phenomena resulted in a paradigm shift in quantum physics during the latter half of the twentieth century. Such collective phenomena can be observed in quantum many-body systems, spanning the disciplines of condensed matter, atomic and molecular physics, and quantum optics. Examples include superradiance and superconductivity, among others \cite{BRANDES2005315,gross_superradiance_1982, RevModPhys.64.313}.  

In his seminal 1954 paper, Dicke introduced the term `superradiance' for the collective, coherent spontaneous emission from a many-atom system. He showed that the superradiance would occur when a collection of excited atoms radiate due to interactions with the common radiation field \cite{dicke_coherence_1954}. From a semi-classical approach, he showed that the intensity of the emitted spontaneous radiation in these systems exhibits a sharp burst that decays much more rapidly than the monotonous 
exponential decay observed in a single excited atom. By changing the inter-atomic distance, a smooth cross-over from the monotonic decay to collective decay was found \cite{gross_superradiance_1982, masson_many-body_2020,masson_universality_2022}. Ever since, attempts have been made to provide a complete quantum mechanical description of this phenomenon \cite{lehmberg_radiation_1970, agarwal_master-equation_1970, rehler_superradiance_1971, bonifacio_quantum_19711, bonifacio_quantum_19712, degiorgio_approximate_1971, haake_quantum_1972}. The first explanation using a simplified dynamical equation in terms of the density matrix of the system to provide a better insight into the microscopic description of the process was proposed in 1970 \cite{agarwal_master-equation_1970} and is known as the superradiance master equation. In the later years, several aspects of the problem, like the statistical properties of the solutions to the dynamical equations and the geometry dependence of the phenomenon, were studied thoroughly \cite{bonifacio_quantum_19711, bonifacio_quantum_19712, haake_quantum_1972,friedberg_temporal_1974}.   

Despite theoretical progress, it took nearly two decades after Dicke's work for the first experimental verification of superradiance to be reported in a system of optically pumped Hydrogen Fluoride molecules \cite{skribanowitz_observation_1973}. This was followed by similar observations in several optical and condensed matter systems using the pulsed dye-laser \cite{crubellier_doppler-free_1978, moi_rydberg-atom_1983, Burlamacchi:76}. Several experimental advances using optical cavities and collection of Rydberg atoms were also reported over the years \cite{doi:10.1126/science.1217901, stiesdal_observation_2020,suarez_superradiance_2022}. In recent years, circuit quantum electrodynamics has enabled the use of artificial atoms and has proven to be an interesting platform to observe such collective effects \cite{mlynek_observation_2014}.

In Dicke's analysis, the interatomic distance was chosen to be much lesser than the wavelength of the radiation field, such that the indistinguishability of the atoms gave rise to collective dissipation on coupling with the field, and all forms of interatomic coupling couplings were ignored. It was also shown that by including symmetric dipolar coupling between the atoms, such collective phenomena remain unchanged. However, for dissimilar dipolar coupling between the atoms, the permutation symmetry is broken, which results in a suppression of such collective phenomena \cite{gross_superradiance_1982}.

Around the same time as Dicke's work on superradiance, Bloembergen showed that spontaneous radiation damping occurs in nuclear spins in the presence of an electric resonant feedback field \cite{bloembergen_radiation_1954}, where the emitted radiation was found to have similarities with the Dicke superradiance phenomena. Despite the differences in the origins of these two processes, such radiation damping is also often called superradiance in a multispin system \cite{benedict1996}. 
As an explanation for this phenomenon, a theoretical analysis using the phenomenological Bloch equation was proposed, which showed that the superradiance in these systems originated from the thermal Nyquist noise of the resonant circuit \cite{bloembergen_radiation_1954}.

However, a strong objection was raised by Yukalov regarding the presence of thermal noise in this problem on the grounds of qualitative mismatch between the theoretical prediction and experimental observations \cite{yukalov_origin_1995,yukalov_nonlinear_1996}. He argued that the origin of superradiance in multispin systems is the presence of local interaction \& not the thermal noise in the circuit. Since the local interaction in NMR is identified as the dipolar interaction, the common environment effect in such cases is negligible. Yukolav proposed that the nonsecular part of the dipolar interaction contributes to the collective relaxation of the system. In this case, such a collective relaxation process was called pure superradiance because, in the absence of initial coherence, the correlation builds up through a purely self-organized process \cite{yukalov_origin_1995}. He provided a set of stochastic equations to explain this process, as the Bloch equation failed to describe such collective phenomena. Since the proposed model was semi-classical, a complete quantum mechanical picture to describe this phenomenon is still an open problem.

Generally, the dynamics of a system weakly coupled with the thermal bath at finite temperature is governed by the Born-Markov quantum master equation (QME) \cite{breuer2002}. The formulation of the QME is based on the independent rate approximation, where any local interaction (e.g., periodic drive, dipolar interaction) appears in the first order, and the dissipation is solely due to the second-order contribution of the system-bath coupling \cite{cohen2004}. For spin-half systems, the Born-Markov QME results in the Bloch equation. Hence, the Bloch equation can only provide the notion of the spin-lattice relaxation process captured by the second-order terms of system-bath coupling. As a result, for a dipolar system, such an equation describes the first-order effects of the secular terms of the interaction and fails to include the second-order terms that come from both the secular and nonsecular parts. Bloembergen showed that the spin-spin relaxation process originates from the second-order contribution of the nonsecular dipolar interaction \cite{bloembergen_relaxation_1948}. In the case of pure spin superradiance, as Yukalov proposed, contributions from the spin-lattice and spin-spin relaxation processes are crucial to the dynamics \cite{yukalov_origin_1995}. Therefore, the problem demands the use of a modified QME capable of capturing both the relaxation processes. 

Recently, Chakrabarti \etal proposed a fluctuation-regulated quantum master equation (FRQME), which can be used for calculating the second-order effects of any local interaction along with the system-bath coupling \cite{chakrabarti2018b}. Later, FRQME was applied to study several phenomena in quantum optics, quantum computation, and information processing \cite{chatterjee_nonlinearity_2020,chanda2020,chanda2021}. In the case of dipolar systems, this formalism can predict the existence of second-order terms that come from the secular and nonsecular dipolar interaction where the real part of the second-order terms results in a Lorentzian absorption line shape  \cite{bloembergen_relaxation_1948}. Therefore, such formalism can explain both the spin-spin and spin-lattice relaxation process using a single dynamical equation. The effect of such terms on the linewidth of the magic angle spinning spectrum (MAS) in NMR, on the entanglement storage device, and the lifetime of the prethermal phase was recently studied by the same author \cite{saha2022,saha_effects_2022, saha_cascaded_2023}.

In this paper, we address the following question: in the case of a dipolar network, can the interactions lead to a collective dissipation in the presence of thermal fluctuations in the local environment instead of coupling via a common environment? We seek the answer by investigating the effects of both the second-order secular and nonsecular terms of the dipolar coupling and system-bath coupling using FRQME. This work strives to provide a dynamical equation similar to the superradiance master equation that explains superradiance in dipolar coupled spins in the presence of local thermal fluctuations. Through this approach, we naturally account for the nonsecular terms arising from the interactions, thereby rigorously proving Yukalov's conjecture \cite{agarwal_master-equation_1970,yukalov_origin_1995}. We also note that this description accounts for environment-driven relaxation along with spontaneous emission \cite{vorrath_dynamics_2005}.

 The organization of the paper is as follows. In section \ref{secii}, we briefly discuss the description of a system of dipolar coupled \(N\) spins. In section \ref{seciii}, we present the dynamical equation for this system using FRQME. We describe the non-equilibrium dynamics of the system under dipolar interaction in section \ref{seciv}. Section \ref{secv} is devoted to the discussion on the comparison of our theoretical analysis with the existing experimental results. Finally, in section \ref{Discussions}, we discuss the results and conclude in section \ref{conclusion}.

\section{Description of the system}
\label{secii}
We consider an ensemble of dipolar coupled N spin-$1/2$ particles, where each spin is weakly coupled with its local thermal environment in the presence of a Zeeman field. The full Hamiltonian for the system can be written as,
\begin{eqnarray}
\mathcal{H} = \hs^{\circ}+\hl^{\circ} + \hdd +  \hsl+ \hl(t)  .
\label{ham1}
\end{eqnarray}
The first term is the free Hamiltonian of the system. Here we define, $\hs^{\circ} = \sum\limits_{i=1}^N \frac{\omega_{\circ}}{2} \sigma_z^i $ where $\omega_{\circ}$ is the Zeeman frequency of the spins and $\sigma_i$ is the Pauli spin matrix $[i = x,y,z]$ . We assume that all the spin-1/2 particles have the same energy levels. The second term, $\hl^{\circ}$, is the free Hamiltonian of the thermal bath. Here, $\hl^{\circ}=\sum_i \omega_L L_{+i}L_{-i}$, where $\omega_L$ is the frequency of the bath and $L_{\pm i}$ stands for the creation and annihilation operators of the bath. The system-environment coupling is defined by $\hsl$ that can be modeled using the Jaynes-Cummings Hamiltonian, $\hsl = \omega_{\rm SL} \left( \sigma_{+i} L_{- i} + h.c \right)$; where $\omega_{\rm SL}$ is the coupling amplitude and $\sigma_\pm$ are the raising and lowering operator corresponding to the $i$th spins. 

$\hdd$ represents the dipolar interaction in the system. We note that in
our case, the nature of the dipolar interaction is different from the Dicke model, where such
interactions are mediated by the photons of  common radiation field
\cite{gross_superradiance_1982}. Here we consider a direct dipolar interactions as
encountered in magnetic resonance spectroscopy
\cite{melinda2001, smithii}. 
Two spins having non-zero magnetic moment in the presence of Zeeman field interact with each
other when they are physically close. 
We also note that the direct interaction between two magnetic dipoles 
has no connection with the common system-environment coupling.   The
analytical form of $\hdd$ is given below.
\begin{eqnarray}
\hdd = \sum_{i,j=1}^N \frac{\mu_{\circ} \hbar}{4
\pi}\left(\frac{\vec{\mu_i}.\vec{\mu_j}}{r^3} -
\frac{3\left(\vec{\mu_i}.\vec{r}\right)\left(\vec{\mu_j}.\vec{r}\right)}{r^5}\right),\quad
\forall \, i >j.
\label{hamdd} 
\end{eqnarray} 
Here, $\vec{\mu} = \gamma \vec{\sigma}/2$,  $\gamma$ denotes the gyro-magnetic ratio, $\mu_{\circ}$ is the magnetic permeability constant and $r$ is the distance between any two spins. The irreducible spherical tensor representation of $\hdd$ is written as, 
\begin{eqnarray}
\hdd = \sum\limits_{i,j=1}^N\sum\limits_{m = -2}^2 \omega_{d_m}(\mathcal{T}_2^m)_{ij}, 
\quad \forall\, i>j.
\end{eqnarray} 
We define, $\omega_{d_m} = \omega_d Y_2^{-m}(\theta, \phi)$ and $\omega_d =
\frac{\hbar \mu_{\circ}\gamma^2}{4 \pi r^3} $ where, $Y_2^{-m}$ is the spherical
Harmonics. $\theta$ and $\phi$ are, respectively, the polar and azimuthal angles between the
orientation of the dipolar vector with the strong Zeeman field. $\mathcal{T}_2^m$ is the
irreducible spherical tensor of rank 2. The definitions of `secular' and `nonsecular' Hamiltonians are borrowed from NMR  literature\cite{melinda2001}. The secular part commutes with the Zeeman Hamiltonian, and the nonsecular part does not commute with the Zeeman Hamiltonian. As such, secular parts do not pick up an oscillating time dependence in the interaction representation, whereas the nonsecular part contains a time-dependent part. The $m=0$ component represent the secular part of $\hdd$ and $m=\{\pm1,\pm2\}$ stand for the non-secular components \cite{saha_effects_2022}.

To obtain the maximum intensity of the radiation burst in the case of the Dicke superradiance problem, the permutation symmetry of the atoms is assumed. As a result, the dynamics can be solved using the collective basis, which keeps the Hilbert space dimension finite. Hence, it is a theoretical simplification for understanding the complicated many-body quantum dynamics. Later, the dependence of radiative intensity on the geometry of the arrangement
of atoms was also studied by several authors \cite{gross_superradiance_1982}.
Motivated by the above approaches, to obtain the maximum correlation created by the nonsecular dipolar correlation, we also assume the atoms are indistinguishable. Furthermore, we have also modeled the dipolar interaction as a mean dipolar interaction (i.e., $\omega_d$ is the mean dipolar coupling amplitude) between the spin-pairs. We note that such an assumption was already being used in solving several problems in dipolar systems \cite{Kucsko2018, peng_floquet_2021, beatrez_floquet_2021, beatrez_critical_2023, Rovny2018, Ho2017, choi_observation_2017}. It is an ansatz that helps us to solve the dynamics using the collective basis and reduce the dimensionality issue in this problem, as we will see from our
analysis in the next sections. On the other hand, how $r_{ij}$, $\theta_{ij}$ and $\phi_{ij}$ affect the correlation for a particular configuration of the dipolar network is not explored here and remains an open problem.

 $\hl(t)$ represents the explicit presence of thermal fluctuations in the local environment, whose form is given by $\hl(t) = \sum\limits_i f_i(t) \vert \xi_i \rangle \langle \xi_i \vert$ such that it is diagonal in the basis of $\hl^{\circ}$. $f_i(t)$ is assumed to be  $\delta$ correlated Gaussian white noise with a second moment $\kappa$. $\overline{f_i(t)f_j(t-\tau)} =\kappa^2  \delta_{ij}(\tau)$. The fluctuations ensure that the bath coherences decay exponentially with a characteristic time constant $\tau_c$, where $\tau_c = 2/\kappa^2 $. The existence of such fluctuations can destroy the coherences in the local environment within a timescale $\tau_c$, but the diagonal elements of the thermal density matrix remain unchanged. Therefore, the fluctuations do not destroy the equilibrium of the bath.

A clear separation of timescale exists in the system, which is given by $\tau_c \ll \Delta t \ll t_s$. Here, $\Delta t$ is defined as the coarse-grained time scale as introduced by Cohen-Tannoudji \etal \cite{cohen2004}, and $t_s$ is the system-relaxation time scale. In this coarse-grained time scale, the propagator is constructed in such a way that the whole system evolves infinitesimally under $\hdd + \hsl$, while the bath evolves by a finite amount under $\hl(t)$ \cite{chakrabarti2018b}. The cumulant expansion of the environment fluctuations of the propagator gives rise to a memory kernel $(e^{-\frac{\tau}{\tau_c}})$ \cite{chakrabarti2018b}. Finally, taking partial trace over the environment operator results in a Markovian quantum master equation with a memory kernel in second-order terms, known as FRQME, that provides a dynamical equation for the system \cite{chakrabarti2018b}. In the interaction picture of $\hs^{\circ} + \hl^{\circ}$, the FRQME is written as,

\begin{eqnarray} \label{frqme1}
\frac{d\rh (t)}{dt}&=& -i \Tr_{\rm L}\Big[\heff(t),\rh(t) \otimes\rl\Big]^{\rm sec}\nn\\
&&-\int\limits^{\infty}_0 d\tau \Tr_{\rm L}\Big[\heff(t),\Big[\heff(t-\tau),\rh (t) \otimes\rl\Big]\Big]^{\rm
sec}e^{-\frac{\tau }{\tau_c}},
\end{eqnarray}
Here, $\heff(t)=\hsl^I +\hdd^I$. `$I$' denotes the interaction representation \emph{w.r.t} free Hamiltonian of the system $(\hs^{\circ}+\hl^{\circ})$. $\rl$ is the
equilibrium density matrix of the environment, and `sec' denotes the secular
approximation \cite{cohen2004}. The presence of an exponential kernel
$\left(\exp(-t/\tau_c)\right)$ results in a finite second-order contribution of $\hdd$
along with $\hsl$. The above dynamical equation (Eq. (\ref{frqme1})) can be reduced to the
Gorini-Kossakowski-Sudarsan-Lindblad (GKSL) form; so CPTP (complete positivity and
trace preservation) holds. A detailed derivation of the FRQME (Eq. \ref{frqme1}) is provided in appendix \ref{appendix-1}.

\subsection{Applicability of FRQME in dissipative dipolar systems} 
 In the case of a dissipative dipolar system, both the spin-spin and spin-lattice relaxation processes are important \cite{yukalov_origin_1995}. For liquids and gases, where high motional narrowing occurs, we note that  $\overline{\omega_{dm}} = 0$. Hence no first-order terms coming from $\hdd$ contribute to the dynamics. The polar and azimuthal angles $\{\theta, \phi\}$  randomly vary in time due to molecular reorientation. As a result. The correlation between the $Y_2^{\pm m} (\theta, \phi)$ [as given in dipolar Hamiltonian] decays exponentially in time. To calculate the analytical expression, one needs to average  $\{\theta, \phi\}$ over a solid angle $4 \pi$. In our theoretical prescription, the explicit presence
of thermal fluctuation successfully captures similar second-order effects of interaction that
has been predicted by previous approaches \cite{bloembergen_relaxation_1948}.
 
 On the other hand, in the case of a liquid being slowly frozen to approach a solid where the dipolar vector moves in a smaller solid angle along the average dipole moment of the spin-pairs, the first-order terms coming from $\hdd$ has a finite contribution in the dynamics. A case in point is liquid crystal, where one experiences residual dipolar coupling (average of the dipolar vector over an incomplete spherical surface) and also has dipolar relaxation \cite{bhattacharyya2003}. In case of further freezing (i.e., for solids), the dipolar vector moves within a very narrow solid angle. Therefore, the time-dependent perturbation theory is still applied in this case, and an exact result can be obtained by averaging over all choices of the average dipole moment of the spin-pairs. For simplicity, we assume that the dipole moments are identical. We also note that the environmental correlation time $(\tau_c$) plays a vital role. In the case of liquids, the value of $\tau_c$ is small, whereas, for solids, the value of $(\tau_c)$ is relatively large \cite{bloembergen_relaxation_1948}. The Redfield equation and  Bloembergen's approach are unable to capture such effects in solids \cite{bloembergen_relaxation_1948, redfield1957}. Similarly, the Born-Markov master equation only describes the spin-lattice relaxation process, and it fails to explain the spin-spin relaxation process \cite{cohen2004,breuer2002}. Therefore, we find that FRQME is useful for such cases \cite{chakrabarti2018b}.
 
 Using FRQME, we have recently successfully explained the emergence of prethermal phases and discrete time crystals in dipolar solids \cite{saha_cascaded_2023, saha2024} in the presence of periodic drive, where the second-order dipolar contributions play a pivotal role in the dynamics. Our analysis is in excellent agreement with the existing experimental evidence \cite{beatrez_floquet_2021,beatrez_critical_2023}. Recently Harkins \etal used Nakazima-Zwanzig formalism to obtain the second-order contribution of the dipolar interaction between $^{13}$C atom and NV center to explain the dynamical stabilization in diamonds \cite{harkins_nanoscale_2023}. In our case,  a similar expansion scheme (i.e., FRQME) is used to observe the effects of second-order atom-atom dipolar interactions. 
 
 We further ensure that no such double counting occurs in such cases while calculating the second-order effects. For example, in the case of dipolar solids, the first-order terms provide an energy shift, which results in a Fourier peak in the frequency domain, while the second-order terms give the spectrum linewidth.

For the remaining part of the manuscript, we will not confine ourselves to the particular case of liquids or gases and provide a generic description of the dipolar systems, as FRQME is applicable for both the liquid state and solid state systems.

\section{Dynamics of the system}
\label{seciii}
In this section, we describe the dynamics of the system using FRQME. The dynamical equation for the $N$-spin density matrix ($\rho_s^N$) using Eq. (\ref{frqme1}) is given by,
\begin{eqnarray}
\frac{d \rh^N}{dt} &=& \mathcal{D}_{d}[\rh^N ] + \mathcal{D}_{\scriptscriptstyle \rm SL}[\rh^N ]
\label{qme-1}
\end{eqnarray}
The term $\mathcal{D}_{d}[\rh^N ]$ stems from the first-order and second order contribution of $\hdd$. The explicit form of $\mathcal{D}_{d}[\rh^N ]$ is given as,
\begin{eqnarray}
\mathcal{D}_{d}[\rh^N ] &=& \sum_{i,j,k,l=1}^{N} \sum_{m=0}^{2}\Big(-i \left[\left(\omega_{d_0} \mathcal{T}_2^0\right)_{ij}  , \rh^N \right]\nn\\
&& - \Gamma(m)\left[(\omega_{d_{\pm m }} \mathcal{T}_2^{\pm m})_{ij},\left[(\omega_{d_{\mp m }} \mathcal{T}_2^{\mp m})_{kl}, \rh^N \right]\right]\Big) \qquad[\forall i >j,k>l]. 
\label{qme-d}
\end{eqnarray}
The real part of the second order contribution of $\hdd$ signifies the spin-spin relaxation process.
Here, $ \Gamma(m) = \frac{\vert\omega_{d_m}\vert^2\tau_c}{1 + (m\omega_{\circ}
\tau_c)^2}$, is the dipolar relaxation rate. We neglect the contribution of the
second-order shift terms of nonsecular interactions in Eq. (\ref{qme-1}).
$\mathcal{D}_{\scriptscriptstyle \rm SL}[\rh^N ]$, signifies the contribution from $\hsl$
(spin-lattice relaxation process) which is written as,
\begin{eqnarray}
\mathcal{D}_{\scriptscriptstyle \rm SL}[\rh^N ]  &=&  \sum\limits_{j=1}^N  p_{\mp}\left[  2\sigma_{j\mp} \rh^N \sigma_{j\pm}  - \{\sigma_{j\pm} \sigma_{j\mp} , \rh^N\}   \right]. 
\label{lindblad}
\end{eqnarray}

In the above case, we do not consider a common environment because, for nuclear spins, dipolar coupling does not require a common environment \cite{yukalov_nonlinear_2005}. Two magnetic dipoles can interact with each other through their magnetic field \cite{melinda2001}. As such, while the nuclear spins relax due to the local environment, they can be dipolar coupled to their neighbors. In addition, we account for the fluctuations in the local environment, which contribute to the dipolar relaxation \cite{bloembergen_relaxation_1948}. 

Here, the effect of the Lamb shift is also neglected. $p_-, p_+$ are respectively defined as the downward and upward transition rate due to $\hsl$, which can be calculated from the spectral density function for the system-environment coupling. Their form is given by
\begin{eqnarray}
p_\pm  = \rm{Re}\left( \int\limits_0^\infty d\tau   \left[\omega_{\rm SL}^2 e^{-\tau/\tau_c} e^{\mp i  (\omega_L - \omega_{\circ})\tau}\Tr_{\rm L} \{L_\pm L_\mp \rl\}\right]\right).
\label{spec}
\end{eqnarray}
$\rm{Re}$ denotes the real part of the expression. The relaxation time $\tau_1$ is defined as $\tau_1 = 1/(p_+ + p_-)$. In the absence of $\hdd$, corresponding equilibrium magnetization, $M_z^{\rm eq}$, is written as, $M_z^{\rm eq} = \frac{p_+ - p_-}{p_+ + p_-}$.

\begin{figure*}[htb]
\raisebox{3cm}{\normalsize{\textbf{(a)}}}\hspace*{-1mm}
\includegraphics[width=0.4\linewidth]{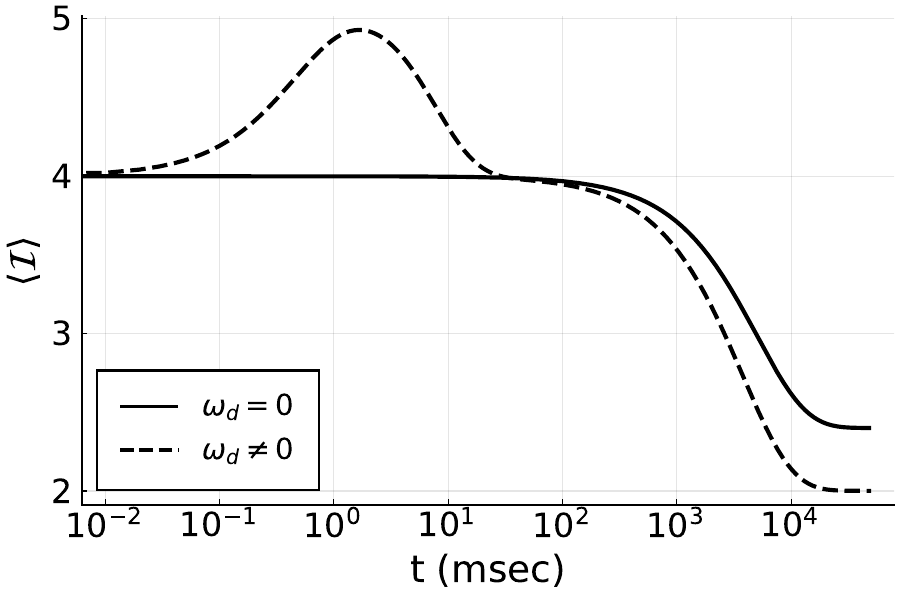}
\raisebox{3cm}{\normalsize{\textbf{(b)}}}\hspace*{-1mm}
\includegraphics[width=0.4\linewidth]{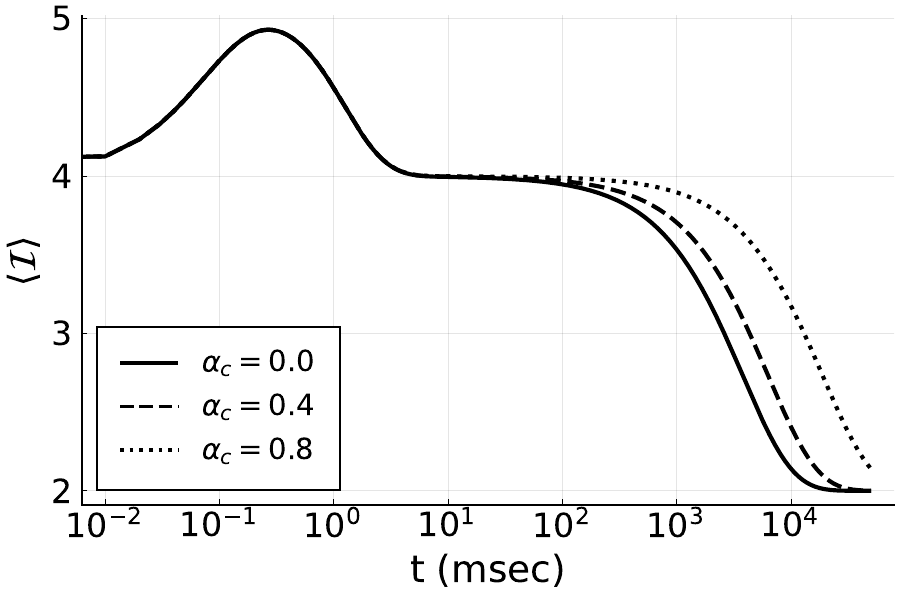}
\caption{The figure shows the full dynamical evolution of radiative intensity ($\langle
I(t)\rangle$) as a function of time for a four-spin dipolar coupled system.
Initially, all the spins are excited. For Fig. (a),  $\omega_d = 2  $ MHz, $\tau_c = 0.1$
$\mu$sec, $\omega_{\circ} = 2\pi \times 100$ MHz, $\theta = \frac{\pi}{4}$,  $p_+ \approx
6 \times 10^{-5}$, $p_-  \approx 4 \times  10^{-5}$. The black line shows the dynamics for
$\omega_d = 0$, and the decay profile is monotonic. On the other hand, the black dashed
line shows the dynamics for the case where $\hdd$ is included. The radiative intensity
gets a sharp peak and then decays at a longer time scale. For Fig. (b), we consider a spatially correlated bath instead of an uncorrelated bath. Here, $\omega_d = 5 $ MHz, while all the other parameters remain the same. Our result indicates that in the limit $\omega_d \gg \omega_{\rm SL}$, for increasing the spatial correlation, the dynamics in the inter-mediate scale remain the same. However, the system reaches the final steady state at a longer time.
 }
\label{fig-1}
\end{figure*}

The above Eq. (\ref{qme-1}) can also be written in the Liouville space as $\frac{d\hat{\rho}_s^N}{dt}= \hat{\hat{\mathcal{L}}}{\hat{\rho}}_s^N$, where $\hat{\hat{\mathcal{L}}}$ is the Liouvillian super-operator. For $N$ spin $1/2$ system, the dimension of the Hamiltonian is  $2^N$ and the dimension of $\hat{\hat{\mathcal{L}}}$ is $2^{2N}$. Therefore, increasing the number of spins in the system makes the dynamical equation of $\rho_s^N$ harder to simulate numerically.   In order to study superradiance, we look at the operators corresponding to the radiative intensity ($I$), defined as,
\begin{eqnarray}
I = J_+ J_- 
\end{eqnarray} 
where, $J_{\pm } = \left( \sum\limits_{i=1}^N \sigma_{\pm i} \right)$.
In Fig. \ref{fig-1}, we plot $\langle I(t) \rangle$ as a function of time for two
choices of $\hdd$, (i) $\omega_d = 0$, and (ii) $\omega_d > \omega_{\rm SL}$ for the four spin
system. The initial state is chosen to be the same as Dicke's superradiance problem,
$\vert \psi \rangle \stoo= \vert \uparrow \uparrow .. \uparrow_N\rangle $
\cite{dicke_coherence_1954}. For such a choice of the initial state, the first-order contribution due to $\hdd$ is zero, as the above-mentioned state is the eigenstate of the secular part of $\hdd$. Therefore, in the case of both liquids and solids, the main contribution comes from the second-order effects. For case (i), where the dipolar coupling is set to zero, the time evolution shows monotonic decay. On the other hand, for case (ii), where the dipolar coupling is set to a non-zero value, an increment in $I(t)$ is observed in an intermediate time scale where the dipolar coupling is much stronger than the system-bath coupling. Therefore, we note that this increment of $I(t)$, which is defined as a radiative burst, is associated with the dipolar interaction. As the initial state $\vert \psi \rangle \stoo$ is the eigenstate of the secular part of $\hdd$, only the nonsecular part contributes to the dynamics. 

We also check the effect of the spatial correlation between the different local environments in the radiative intensity. In such a case, where the spatial correlation is taken into account, the cross terms between different spin-bath coupling become effective \cite{saha_effects_2022}. For those cross terms, $p_-, p_+$ are further modified by an $\alpha_c$ factor, where $\alpha_c \propto e^{-r/\xi_c}$, such that $p_{\pm}$ is replaced by $ \alpha_c p_{\pm}$ \cite{Jeske2013}. Here, $r$ is the distance between the spin pairs, and $\xi_c$ is the bath-correlation length. If $r \to 0$ or $\xi_c \to \infty$, the environment acts as a common environment $(\alpha_c \to 1)$, which is an asymptotic limit and is not applicable for our case. For, $r \gg \xi_c$, the environment acts as a local environment $(\alpha_c =0)$. In the limit, $\omega_d \gg\omega_{\rm SL}$, the dynamics in the intermediate timescale have no dependence on the spatial correlation between the different local environments as the dynamics are dominated by $\hdd$ in this regime. For increasing spatial correlations, the system spends more time in the quasi-steady state and reaches the final equilibrium at a longer time, which we have shown in your earlier work \cite{saha_prethermalization_2024}. As our main focus is to describe the underlying physics behind the radiative burst in the intermediate time scale, we will use the local environment with no spatial correlation $(\alpha =0)$ for simplicity. 
 
\begin{figure*}[htb]
 \raisebox{3cm}{\normalsize{\textbf{(a)}}}\hspace*{-1mm}
\includegraphics[width=0.4\linewidth]{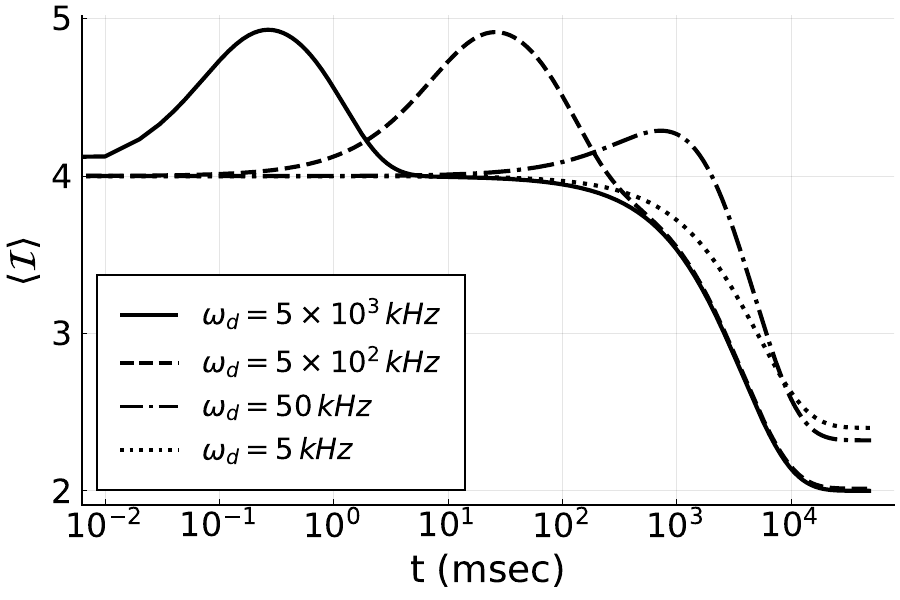}
 \raisebox{3cm}{\normalsize{\textbf{(b)}}}\hspace*{-1mm}
\includegraphics[width=0.4\linewidth]{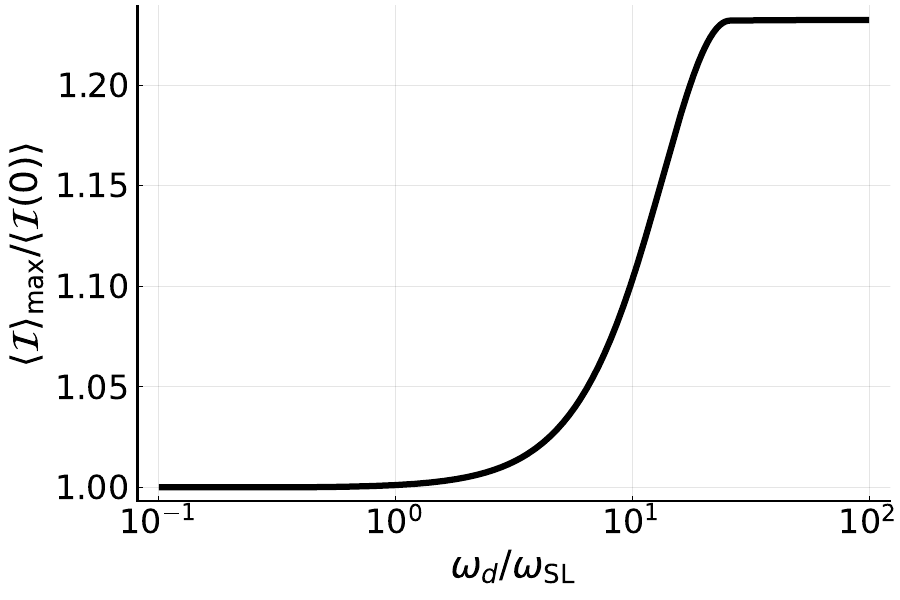}
 \raisebox{3cm}{\normalsize{\textbf{(c)}}}\hspace*{-1mm}
\includegraphics[width=0.4\linewidth]{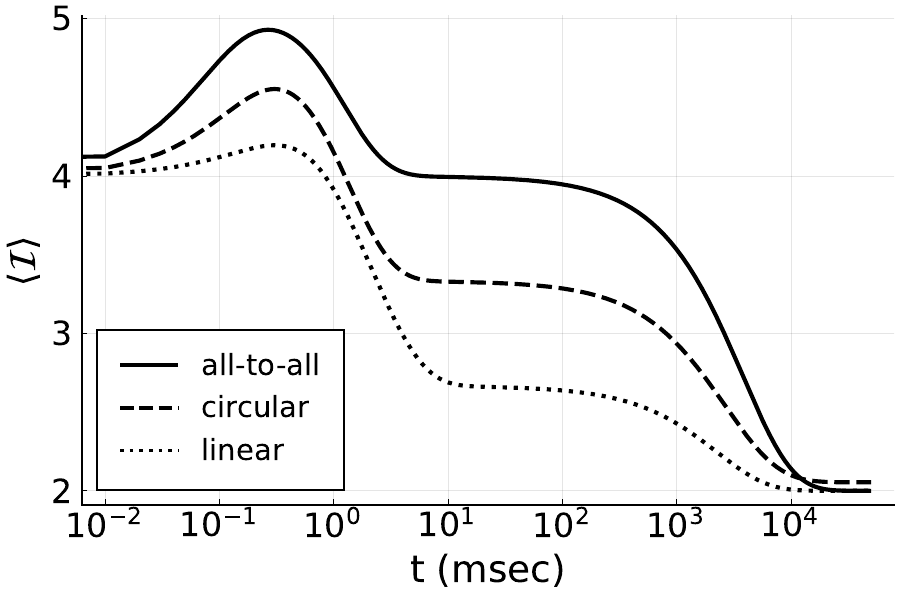}
\raisebox{3cm}{\normalsize{\textbf{(d)}}}\hspace*{-1mm}
\includegraphics[width=0.4\linewidth]{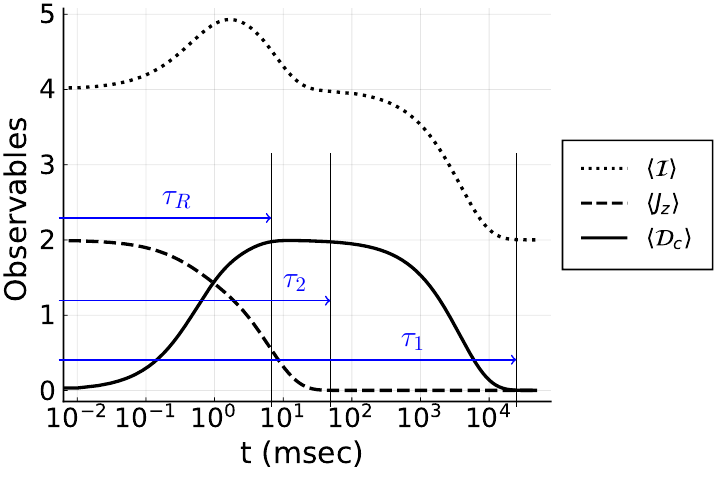}
\caption{   The initial state is chosen as $\vert \psi \rangle \stoo =
\vert \uparrow \uparrow \uparrow \uparrow \rangle$ for all the above plots.
 We assume that, for Fig. (a), (b) and (d), all the spin-pairs are dipolar coupled to each other. Fig. (a) shows the plot of $\langle I(t) \rangle$ vs time for four different choices of dipolar interaction amplitude. The values of the fixed parameters are given as, $N=4$, $\omega_\circ = 2 \pi \times 100$ MHz, $\tau_c = 0.1$ $\mu$sec, $\theta = \frac{\pi}{4}$,  $p_+ \approx 6 \times 10^{-5}$, and  $p_- \approx 4 \times  \times 10^{-5}$. When $\omega_d >\omega_{\rm SL}$, a radiation burst is obtained. By decreasing the dipolar coupling amplitude, such a peak arises at a later time. For the opposite limit, $\omega_d <\omega_{\rm SL}$, only a monotonic decay is observed. In Fig (b),  maximum intensity $(I_{\rm max} / I(0))$ is plotted as a function of $\omega_d/ \omega_{\rm SL}$, which shows that the radiation burst occurs for $\omega_d >\omega_{\rm SL}$ and remains constant beyond a threshold value of $\omega_d/ \omega_{\rm SL}$. Fig. (c) shows the plot of $\langle I(t) \rangle$ vs. time for different choices of the geometry of the system. The maximum burst is possible in the presence of all-to-all dipolar coupled systems. For the circular case, it becomes lower than the previous case and is the least for the linear case. Here $\omega_d = 5  $ MHz.   Fig. (d) shows the plot of the observables $\langle I(t) \rangle$, $\langle \mathcal{D}_c\rangle$, and $\langle J_z \rangle$ as a function of time and several existing timescales present in the system is also shown. Here $\omega_d = 2  $ MHz. $\langle \mathcal{D}_c\rangle$ grows to a quasi-steady state at a time scale $\tau_R$. Similarly, $\langle J_z\rangle$ decays at a time-scale $\tau_2$. Finally, the system reaches the equilibrium at a time $\tau_1$. Such radiative burst occurs at a time shorter than $\tau_2$.  }
\label{fig-2q}
\end{figure*}

We also note that the peak associated with the superradiance depends on the inter-atomic separation between atoms. More specifically, when the distance between the atoms decreases, the dipolar coupling increases. To illustrate this, we plot $\langle I(t)\rangle$ as a function of time for different choices of $\omega_d$ in Fig. \ref{fig-2q}(a). 
When $\omega_d \gg \omega_{\rm SL}$, such a burst appears at a much earlier timescale. By decreasing the amplitude of the dipolar interaction, the peak appears at a comparatively later time. For the case, $\omega_d \leq \omega_{\rm SL}$, no such peak arises, and the dynamics exhibit monotonic decay. In experiments, instead of one radiation burst, multiple bursts at different times can also be observed. The origin of these multiple peaks can be described using Fig. \ref{fig-2q}(a) by noting that such phenomena can occur in a collection of spins where dipolar interaction between several spin pairs is not identical. For stronger coupled spin-pairs, such a burst is observed at earlier times, whereas, for pairs with a weaker coupling, the burst occurs at comparatively later times.

We also plot the maximum intensity by varying $\omega_d/\omega_{\rm SL}$ in Fig. \ref{fig-2q}(b). The plot shows that the radiation burst happens when $\omega_d >\omega_{\rm SL}$ and that on increasing $\omega_d$ beyond a threshold value, the maximum intensity doesn't increase. Further, to understand how the geometry of the system affects the radiation burst, we also plot $\langle I(t) \rangle$ versus time for different configurations of the system in Fig. \ref{fig-2q}(c). The configurations we study are given below: (i) a dipolar network with all-to-all coupling, (ii) a circular chain of dipolar coupled spins, and (ii) a linear chain of dipolar coupled spins. We observe that the peak is maximum when an all-to-all coupling is present in the system. For circular chains, this peak is lower, and it is the lowest for linear arrangement. We will describe the reason behind such geometry-dependent phenomena later in the manuscript. As the peak is maximum for the all-to-all coupled case, we only consider all-to-all coupled dipolar networks in the remaining part of our analysis. 

The operator corresponding to radiative intensity can be written as a sum of the collective `z' magnetization $(J_z)$ and the dipolar correlation matrix $(\mathcal{D}_c)$. For an N-spin system, it is written as
\begin{eqnarray}
I(t) = \frac{N}{2} \mathds{1}+ J_z + 2\mathcal{D}_c.
\label{intensity}
\end{eqnarray}
Here, $\mathds{1}$ is the identity matrix and $J_z = \sum_i \sigma_{zi}$. The dipolar correlation matrix is given by,
\begin{eqnarray}
\mathcal{D}_c = \frac{1}{4}  \sum\limits_{i,j=1}^{N} (\sigma_{i+} \otimes \sigma_{j-} + \sigma{_i-} \otimes \sigma_{j+})\quad [\forall i > j]
\end{eqnarray}
$\mathcal{D}_c$ is an off-diagonal matrix representing the system's spin-spin correlation. We plot $\langle J_z \rangle$ and $\langle \mathcal{D}_c \rangle$ as a function of time for a dipolar network consisting of four spins in Fig. \ref{fig-2q}(d). The plot shows that three timescales exist in the system. $\langle \mathcal{D}_c \rangle$ grows in a time-scale $\tau_R$ and $\langle J_z \rangle$ decays in a time-scale $\tau_2$. Finally, $\hsl$ provides a very long decay time $\tau_1$. So, we have $\tau_R < \tau_2 \ll \tau_1$. We find that $\langle \mathcal{D}_c \rangle$ grows in a much shorter timescale than the decay time of $\langle J_z \rangle$, which results in a peak of the radiative intensity $(\langle I(t) \rangle)$ in the intermediate timescale. However, the radiative burst does not survive after the time $\tau_2$.
To study the collective properties of the dynamics, we only focus on the time-evolution up to the time $\tau_2$. Therefore, the contribution from $\hsl$ is neglected for our further analysis. 

In the intermediate timescale (up to $\tau_2$), considering all-to-all dipolar coupling, there exists a quasi-conserved quantity in the system, given by, 
\begin{eqnarray}
\frac{d \langle\mathbf{J}^2 \rangle}{dt} = 0
\label{conserved}
\end{eqnarray}
 Here, the conserved quantity, $\mathbf{J}=  \sum_i \mathbf{\sigma_i}/2$ is known as the total angular momentum operator. Following the definition of the symmetry operator in QME,  an operator $\mathcal{O}$ is said to be a symmetry if the corresponding super-operator commutes with the Liouvillian $(\hat{\hat{\mathcal{L}}})$ i.e.,  $\left[\hat{\hat{\mathcal{O}}},\hat{\hat{\mathcal{L}}} \right] = 0$ \cite{buca_note_2012}. For this case, we have $[\hat{\hat{\mathcal{D}}}_d, \hat{\hat{\mathbf{J}}}]=0$, since the effect of $\hat{\hat{\mathcal{D}}}_{\rm \scriptscriptstyle SL}$ is neglected. We note that the Liouvillian super-operator is invariant under the following symmetry transformation $\hat{\hat{\mathcal{U}}} \hat{\hat{\mathcal{D}}}_d \hat{\hat{\mathcal{U}}}^\dagger = \hat{\hat{\mathcal{D}}}_d $ (here $\hat{\hat{\mathcal{U}}} = \exp(-i \Phi \hat{\hat{\mathbf{J}}} )$,  and $\Phi$ is a real parameter). Therefore, we follow the common practice and adopt the collective basis (or angular momentum basis) approach as the particular Hilbert space in this basis grows linearly with increasing the number of atoms \cite{vorrath_dynamics_2005}.

\section{Dynamics under dipolar interaction in the collective basis for $\omega_d \gg \omega_{\rm SL}$}
\label{seciv}

The most important feature of the presence of the symmetry operator $(\mathbf{J}^2)$ is that the dynamics are confined in a particular $\vert J M \rangle$ block. For example, in this case, the excited state is chosen as the initial state of the individual atoms. In the $\vert J M \rangle$ basis, it is written as,  $\vert \psi \rangle \stoo= \vert J=\frac{N}{2},M=\frac{N}{2} \rangle$. Hence, the dynamics is confined to the principal $\vert \frac{N}{2}, M \rangle$ block \cite{vorrath_dynamics_2005}. As a consequence of the presence of such a symmetry operator,  the dipolar interaction can also be written in terms of the collective operators $(J_+,\, J_-$, and $J_z$ ). The form of $\hdd$ using $J_i$ operators is given by,
\begin{eqnarray}
\hdd^{0}  &=& \omega_{d_0}(3 J_z^2 - \mathbf{J}^2)\nn\\
\hdd^{+1} &=& \omega_{d_1}(J_zJ_{+} - \frac{J_+}{2}) , \quad \hdd^{-1} = (\hdd^{+1})^*\nn\\
\hdd^{+2} &=& \frac{\omega_{d_2}}{2}J_+J_{+} , \quad \hdd^{-2} = (\hdd^{+2})^*
\end{eqnarray} 
The matrix element of the collective operators in the $\vert J M \rangle$ basis are given below,
\begin{eqnarray}
J_z &=& M \delta_{ J^{\prime},J} \delta_{ M^{\prime},M}\nn\\
J_{\pm} &=& \sqrt{(J\mp M)(J \pm M+1)} \delta_{ J^{\prime},J} \delta_{ M^{\prime},M\pm 1}
\end{eqnarray}
The initial state of the system is the eigenstate of $\hdd^0$. So, the secular part will not contribute to the dynamics. The dynamics is now confined to $(N+1) \times (N+1)$ dimensions instead of $2^N \times 2^N$. The possible choice of observables in this basis is given by,
\begin{eqnarray}
P^J_{MM^{\prime}}(t) &=& \Tr_{s} \{\vert J M\rangle \langle J M^{\prime} \vert \rho_s^N(t)\}.
\label{observables}
\end{eqnarray} 
 The total number of observables is $(N+1)^2 - 1$ since a constraint of trace preservation exists. Using Eq. (\ref{qme-1}), the dynamical equation of $\rh^N(t)$ in terms of observables is written as,

\begin{eqnarray}
\frac{d}{dt} P^J_{M,M} &=& -\alpha (M) (P^J_{M,M} - P^J_{M-1,M-1}) -\beta (M) (P^J_{M,M} - P^J_{M+1,M+1})\nn\\
&& -\gamma (M) (P^J_{M,M} - P^J_{M-2,M-2}) -\delta (M) (P^J_{M,M} - P^J_{M+2,M+2}) 
\label{qme-2}
\end{eqnarray}
In the Eq. (\ref{qme-2}), $M$ can be varied from $-J$ to $J$.
Here, only diagonal elements contribute to the dynamics, so the number of independent observables is further reduced to $N$. In the next analysis, we simply denote $P^J_{MM}$ as $P_M$. 
The expression for the above rates is obtained to be,
\begin{eqnarray}
\label{rate-1}
\alpha(M) &=& 2\Gamma(1)(J+M)(J-M+1)(M-\frac{1}{2})^2,\nn\\
\beta(M) &=& \alpha(-M),\nn\\
\gamma(M) &=& \frac{\Gamma(2)}{2}(J+M)(J-M+1)(J+M-1)(J-M+2),\nn\\
\delta(M) &=& \gamma(-M).
\end{eqnarray}  
Here, $\alpha (M)$ and $\beta(M)$ contains $\Gamma(1)$ and $\gamma(M)$ and $\delta(M)$ contains $\Gamma(2)$. We note that $\alpha (M)$, $\beta(M)$,  $\gamma(M)$, and $\delta(M)$ are non-linear functions of $M$, which results in the non monotonic decay profile.  $\alpha (M)$ and $\beta(M)$ are the transition rates that arise from the nonsecular parts of the dipolar interaction of rank 1. They are responsible for $\pm 1$ transition, as in the rate equation $P_M$ is connected to $P_{M \pm 1}$ by $\alpha (M)$ and $\beta(M)$. Similarly, $\gamma(M)$ and $\delta(M)$   are responsible for $\pm 2$ transition.

We numerically simulate the above dynamical equation [Eq. (\ref{qme-2})] in terms of $P_M$. The initial condition is chosen as $P_{M=\frac{N}{2}}\stoo=1$. In terms of $P_M$, the expectation value of radiative intensity $(\langle I(t) \rangle)$ is expressed as,
\begin{eqnarray}
\label{eqn-inten}
\langle I(t) \rangle = \sum\limits_{M=-J}^{J}(J+M)(J-M+1)P_M(t)
\end{eqnarray} 
\begin{figure*}[htb]
 \raisebox{3cm}{\normalsize{\textbf{(a)}}}\hspace*{-1mm}
\includegraphics[width=0.4\linewidth]{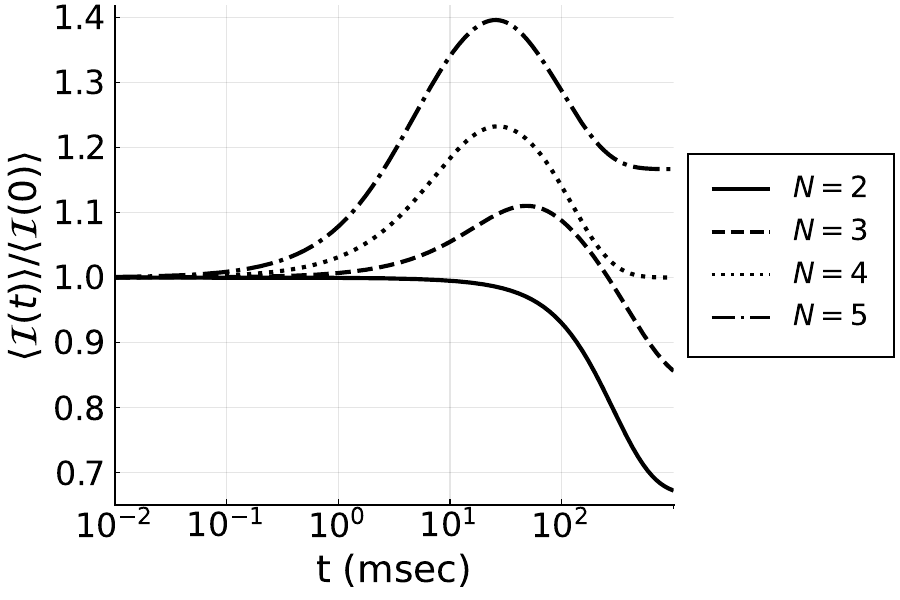}
 \raisebox{3cm}{\normalsize{\textbf{(b)}}}\hspace*{-1mm}
\includegraphics[width=0.4\linewidth]{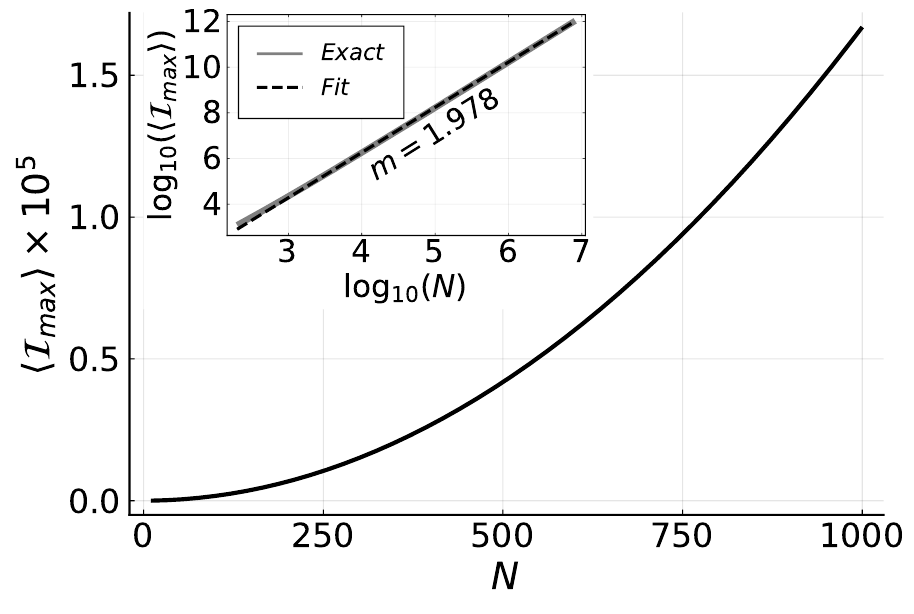}
 \raisebox{3cm}{\normalsize{\textbf{(c)}}}\hspace*{-1mm}
\includegraphics[width=0.4\linewidth]{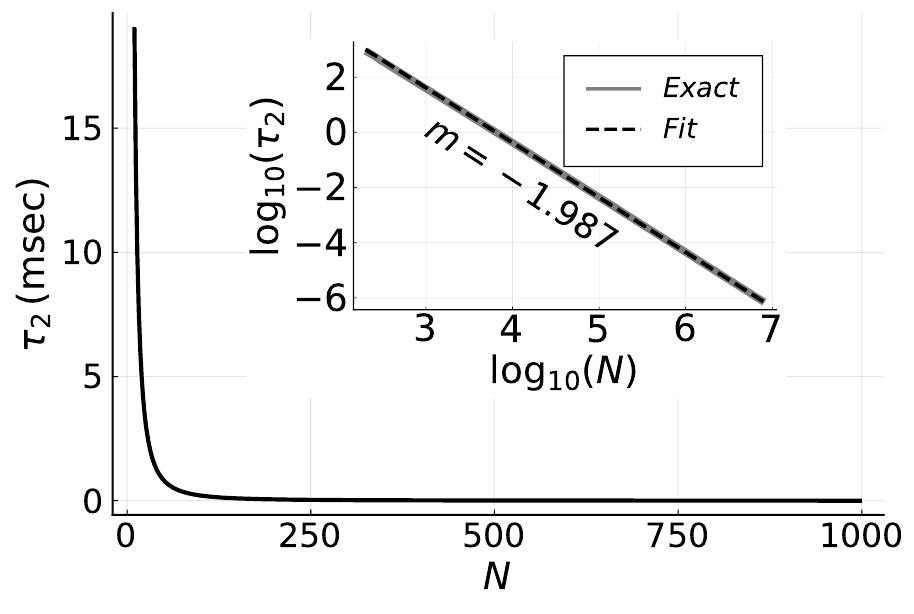}
\raisebox{3cm}{\normalsize{\textbf{(d)}}}\hspace*{-1mm}
\includegraphics[width=0.4\linewidth]{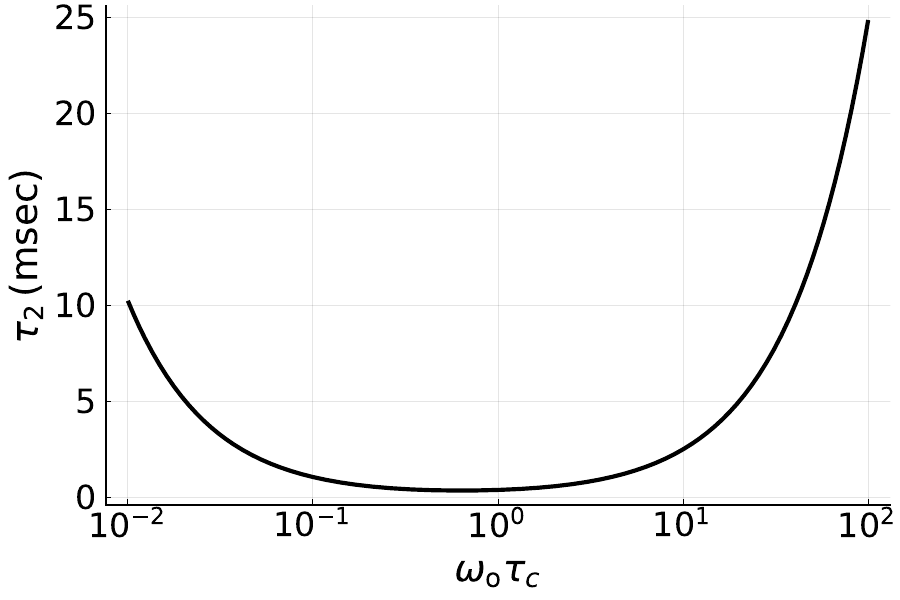}
\caption{Results of the numerical simulations for a dipolar network in the collective basis up to the timescale $\tau_2$ are provided in the following diagrams. The values of the fixed parameters are given as, $\omega_d = 5\times 10^2$ kHz, $\tau_c = 10^{-4}$ msec, $\omega_{\circ} = 2\pi \times 10^5$ kHz, and $\theta = \frac{\pi}{4}$. The initial condition is chosen as $\vert \psi \rangle \stoo = \vert \frac{N}{2}, \frac{N}{2} \rangle $. The time evolution of $\langle I(t) \rangle/\langle I(0)\rangle$ is shown for different atom numbers in (a). For $N=2$, no radiation burst occurs. Such a peak arises for $N>2$. The maximum intensity increases with increasing $N$. Fig. (b) shows the Numerical plots of $\langle I\rangle _{max}$ as a function of atom number ($N$). The number of atoms is varied from ten to thousand. The maximum intensity follows non-linearity in $N$. In the inset, $\log(I_{max})$ is plotted as a function of $\log(N)$ and also fitted with a straight line, $y = m\times N + c$. We find that $m = 1.978,\, c = - 1.649$. Similarly, (c) shows the decay time ($T_{\alpha}$) decreases with $N$, following a non-linearity in $N$. In the inset $\log_{10}(T_{\alpha})$ is plotted as a function of $\log_{10}(N)$ and also fitted with a straight line, where $m= -1.987, \, c= 7.577$. The decay time is measured by calculating the inverse of the ADR of $\hat{\mathcal{D}}_d$.  (d) shows the plot of  $\tau_2$ for varying $\omega_\circ \tau_c$ for $N=4$. The relaxation time is minimum for $\omega_\circ \tau_c = 1$. Here $\omega_{\circ} =  10^5$ kHz and all the other parameters are the same as above.  }
\label{fig-2}
\end{figure*}
We plot $\langle I(t) \rangle$ vs time by changing the number of atoms in Fig. \ref{fig-2}(a). The plot indicates that for a two-spin system, no such radiation burst is observed,  whereas, for $N>3$, the maximum intensity $(\langle I\rangle_{max} )$ increases with $N$. The decay time $(\tau_2)$ shows exactly opposite characteristics to that of $\langle I \rangle_{max}  $ as it decreases on increasing the atom number. Therefore, the radiation burst becomes short-lived and more intense for a higher number of atoms. 

  The maximum intensity $\langle I \rangle_{max} $ is plotted by varying the atom number ($N$) in Fig. \ref{fig-2}(b). The decay time of the system $(\tau_2)$ can be defined as the inverse of the asymptotic decay rate (ADR) of the Liouvillian super-operator $\hat{\hat{\mathcal{D}}}_d$. The definition of ADR is given as the spectral gap between the zero eigenvalues and the first negative eigenvalue of the Liouvillian super-operator \cite{kessler_dissipative_2012}. The decay time is also plotted as a function of the atom number $(N)$ in Fig. \ref{fig-2}(c). Both plots show non-linearity in $N$ (Fig. \ref{fig-2}(b), (c)).
  
We also plot $\langle I \rangle_{max} $ and $\tau_2$ for varying $\tau_c$. The dipolar relaxation rate is minimum for $\tau_c \approx 1/\omega_\circ$. We note that the maximum intensity is nearly constant for varying $\tau_c$. However, at a later time, the dynamics are governed by $\hsl$, and the quasi-conservation law is broken in this regime, therefore enabling the transition between different $\vert J M \rangle$ blocks. As a result, the dynamics of the system show a monotonic decay after the timescale $\tau_2$, and the system reaches the final steady state at a timescale $\tau_1$. 

However, for observing the complete dynamics of the system, one can add
$\hsl$ in the numerical simulation which provides an additional decay of the radiative
intensity at a later time, which does not show any extra new features in the dynamics.

\section{Comparison with the existing experimental results}
\label{secv}

The characteristics of the collective dissipation we observe here have several similarities with the pure spin superradiance in NMR. In nuclear spin systems, the effect of a common electromagnetic field is negligible \cite{yukalov_nonlinear_2005}. In such systems, the spin-lattice relaxation time ($T_1$) is much longer than the spin dephasing rate, which comes from the dipolar interactions ($T_2$), i.e., $T_1 \gg T_2$. For a typical NMR experiment, the spins are initially prepared in an inverted magnetization mode and simultaneously coupled with a resonator. The resonator feedback field in the presence of the nonsecular part of the dipolar interaction plays a pivotal role in the collective relaxation of such a system \cite{yukalov_origin_1995}. One necessary condition required for such a radiation burst is that the radiation time $\tau_\circ$ must be smaller than $T_2$ ($\tau_\circ < T_2$).

The first experimental observations were made by using $^{27}$Al nuclear spins in ruby Al$_2$O$_3$ \cite{bosiger_ruby_1978}. Later, it was observed for protons in C$_4$H$_9$OH \cite{kiselev_discovery_1988}. Recently, magnetic nano-molecules and nano-clusters (which are formed by oxides of Ni, Fe, Co, and Hg) have also been proven as promising candidates to demonstrate such collective phenomena in large spin-systems \cite{yukalov_coherent_2008}. 

In our case (Fig. \ref{fig-2q}(d)), we define $\omega_{\circ} = 2\pi \times 10^5$ kHz and the fluctuation correlation time is $\tau_c = 10^{-4}$ msec. The system-bath coupling amplitude is chosen in such a way that the value of the relaxation time, $\tau_1 \approx 10^4$ msec. Here,  $\omega_d = 2 \times 10^3$ kHz and the corresponding dipolar relaxation time, $\tau_2 \approx 10^2$ msec. In our case, $\tau_2$ plays a similar role to $T_2$. On the other hand, the dipolar correlation appears in a timescale, $\tau_R = 10$ msec. The radiation burst, in our case, arises in a timescale $(\tau_\circ)$,  where $\tau_\circ \approx\tau_R$. We note that this timescale separation $(\tau_\circ<\tau_2)$ emerges naturally in our analysis and is one of our main results.

\section{Discussions}
\label{Discussions}
Collective dissipation can occur in a system of dipolar coupled spins interacting with the local environment. Such phenomena differ significantly from the Dicke superradiance. The latter is due to interaction with the common environment where the dissipation occurs through collective spontaneous emission and requires a minimum of two spins \cite{dicke_coherence_1954}. Also, no direct coupling between them is considered in this problem \cite{dicke_coherence_1954}. On the other hand, a common environment is absent in our description, and the spins are coupled with each other through dipolar interaction. In addition, they are also coupled with their local environment instead of a common environment. The collective dynamics here emerge from the cross terms of the dipolar interactions from different spin pairs and, therefore, require a minimum of three spins. 

We assume that the dipolar interactions between the spins are identical, and along with the spontaneous emission considered in the Dicke model, we also add the effect of absorption and emission due to the thermal environment in our description. For the initially inverted magnetization mode, the secular terms have no contribution. The relaxation process is mainly dominated by the nonsecular part of the interaction in the intermediate regime when the dipolar coupling is much stronger than the system-bath coupling. In order to capture these effects, we use FRQME instead of the usual Born-Markov master equation. We note that the strength of the dipolar interaction increases by decreasing the distance between the atoms. We also find that collective dissipation depends on the density of the system. For dense dipolar networks, when $\omega_d > \omega_{\rm SL}$, the maximum burst occurs, as shown in Fig. \ref{fig-2q}(b). Whereas, for $ \omega_d \approx \omega_{\rm SL} $, the maximum intensity is lower than the previous case, and for $\omega_d < \omega_{\rm SL}$, no such peak is possible, and the dynamics exhibit a monotonic decay.

The dipolar correlations between the different spins build up in a relatively shorter time than the decay of the collective `$z$' magnetization, which results in a radiative burst in the intermediate time. Such correlation in the system builds up due to the cross terms of the interaction between different pairs, whereas the collective `$z$' magnetization decays due to the second-order contribution of the self terms of the coupling. For example, in the case of an $N$ spin system, considering the all-to-all coupling case, the possible number of pairs is $M = {N \choose 2}$. The number of terms contributing to $\mathcal{D}_c$ is ${M \choose 2}$, whereas only $M$ terms are responsible for the decay of $J_z$. We note that, ${M \choose 2}\geq  M$ for $N>2$. It signifies that $\mathcal{D}_c$ grows faster than the decay of $J_z$ when $N\geq  3$. Therefore, at least three dipolar coupled spins are required to observe the radiative burst.

The above analysis also helps us understand the geometry dependence of the system in the
radiation burst. The number of pairs for the linear case is $M_l=N-1$, and for the
circular case, it is $M_c=N$. In case of $N \geq  4$,  we have $M> M_c >M_l$.
Therefore, in the case of a dipolar system consisting of a minimum of four spins, the radiation burst will be maximum when we consider all-to-all coupling between the spins in comparison to the linear and circular cases. This particular observation is in line with the recent work on Dicke superradiance by Masson \etal \cite{masson_universality_2022}. The particle exchange symmetry is preserved for the all-to-all case when the coupling between each pair is equal. Such symmetry leads to the conservation of the total spin operator, $\langle\mathbf{J}^2 \rangle$. Therefore, the all-to-all coupling case is much easier to handle in a collective basis, hence justifying our choice.

The maximum intensity is also plotted as a function of atom number $N$, which shows an
$N^2$ dependence. Therefore, for a higher number of atoms, such a radiative burst is more
intense. Moreover, the dipolar relaxation time $(\tau_2)$ is proportional to $1/N^{2}$,
implying that the radiative burst becomes short-lived on increasing $N$. The short-lived
and intense radiation bursts are the main features of the collective dissipation in the
system. In case of increasing the number of atoms in the dipolar network, we intentionally keep the mean interaction strength the same, such that the system becomes dense. We note that for a dipolar network consisting of $N$ spins, the average interaction strength between the spin pairs is given by $\omega_d \propto \vert \omega_{d_{ij}}\vert/N$, where $\omega_{d_{ij}}$ is the nearest neighbor interaction \cite{lacelle1993}. Therefore, when the number of atoms increases, the average interaction becomes weaker. Here, we keep
it fixed, which is possible when the distance between the atoms is reduced. Such a dense
configuration leads to a more intense and short-lived radiation burst, which matches with
the experimental outcomes \cite{gross_superradiance_1982}.

 The dipolar relaxation rate, $\Gamma(m) \propto \frac{\tau_c}{1 + (m \omega_\circ
\tau_c)^2}$, is  plotted in Fig. \ref{fig-2}(d).  $\alpha (M)$, $\beta(M)$,  $\gamma(M)$, and $\delta(M)$ contains $\Gamma(m)$. For $\tau_c >1 / \omega_\circ $, $\Gamma(m) \propto 1/ \tau_c$ and, for $\tau_c < 1 / \omega_\circ $, $\Gamma(m) \propto \tau_c$. So, in the limit, $\tau_c <1 / \omega_\circ$, we note that for lowering $\tau_c$, the dipolar relaxation time becomes longer. Similarly, in the opposite limit, $\tau_c >1 / \omega_\circ$, the relaxation time further increases for increasing $\tau_c$. Hence for $\omega_\circ \tau_c \approx 1$, the dipolar relaxation time $\tau_2$ becomes the shortest as $\alpha (M)$, $\beta(M)$,  $\gamma(M)$, and $\delta(M)$ are maximum at that point.  Here, we show that, for a dipolar coupled system, in the regime $\omega_d \gg \omega_{\rm SL}$, the lifetime of the radiation burst is minimum at $\tau_c \propto 1 / \omega_\circ$. A short-lived and intense superradiant emission can be used as a superradiance pulse laser \cite{kristensen_subnatural_2023,bohnet_steady-state_2012, laske_pulse_2019, shvets_superradiant_1998, zen_full_2023}. In the case of a fixed Zeeman frequency, one can construct a superradiant laser pulse by choosing the fluctuation correlation time $(\tau_c)$  appropriately. 

Our recent work qualitatively predicts the creation of correlations in a system of dipolar interacting spins. While we have assumed the interaction between spin-pairs to be identical, for a real scenario, the spatial degrees of freedom of the spin-pairs are different and lead to different dynamical signatures. We note that the collective basis cannot be used for analysis in such cases, and therefore, one may face the dimensionality issue. An efficient numerical technique for analyzing such many-body open systems is still an open area of research \cite{Weimer2021}.

\section{Conclusion} 
\label{conclusion}

We have presented a theoretical description of the collective dissipation that arises in a dipolar network
in the presence of thermal fluctuations in the local environment. To this end, we use FRQME, which
successfully predicts the second-order dissipation that comes from local interactions (e.g., dipolar
interaction). We observe that the secular term does not affect the dynamics. On the other hand, the
nonsecular pairs play a predominant role in the origin of collective behavior in the system. In the case of
negligible system-bath coupling, a sudden, short-lived increment in the radiation intensity is observed as
the dipolar correlation builds up for an initially inverted collection of spins; that is, a superradiant
phase appears as an emergent behavior. As the system size is varied, the maximum peak of the radiation
intensity curve scales as $N^2$, while its decay time scales as $1/N^2$. We also show that the time scales
of the radiation burst in our analysis are in good agreement with the experimental results of pure spin
superradiance in NMR.

\section{Acknowledgments}
The authors thank Arnab Chakrabarti and Arpan Chatterjee for helpful
suggestions and insightful comments. SS acknowledges the University Grants Commission for a research fellowship (Student ID: MAY2018- 528071). YI thanks IISER Kolkata for a project grant.

\appendix
\section{ Derivation of FRQME in the context of dissipative dipolar system }
\label{appendix-1}

We note that FRQME was originally reported by Chakrabarti \etal \cite{chakrabarti2018b}. In the paragraphs
below, we provide the key steps leading to the emergence of the master equation.

The total Hamiltonian of the system + environment is given by,
\begin{eqnarray}
\mathcal{H} = \hs^{\circ}+\hl^{\circ} + \hdd +  \hsl+ \hl(t)  .
\label{ham11}
\end{eqnarray}
A discussion about the individual components of the above Hamiltonian (Eq. (\ref{ham11})) is presented in section \ref{secii}. Initially, the system and environment are assumed to be uncorrelated, so $\rho (t) = \rh (t) \otimes \rl$. Here, $\rl$ is the equilibrium density matrix of the system. This approximation is known as the Markov approximation.
 
  In the interaction picture \emph{w.r.t} $\hs^{\circ}+\hl^{\circ}$, the von Neumann Liouville equation of the system + environment is given by,
  \begin{eqnarray}
 \frac{d \rho^I}{dt} = -i[H,\rho^I ]
 \label{vnm1}
 \end{eqnarray}
 Here, $\rho^I(t)$ is the full density matrix of the system in the interaction picture, and $H (t) = \hsl^I(t) + \hdd^I (t)+ \hl^I(t) $. The solution of the above Eq. (\ref{vnm1}), is given as,
 \begin{eqnarray}
 \rho^I(t + \Delta t) - \rho^I(t) = -i \int\limits_t^{t + \Delta t} dt_1 [ H(t_1), \rho(t_1)].
 \end{eqnarray}
 Here, $\rho(t_1) =U(t_1,t) \rho(t) U^\dagger(t_1,t)$, and $U(t_1, t)$ is called the propagator, which has the form,
 \begin{eqnarray}
 U(t_1, t) = T\exp[-i \int_t^{t_1} dt_2 H(t_2)]
 \end{eqnarray}
 where $T$ is the time-ordering operator. To find the dynamical equation of the system, we need to take a partial trace over the environmental operator, such that $\rh(t) = \Tr_{\rm L} \{\rho^I(t)\}$. On taking the partial trace, the above dynamical equation can be written as,
  \begin{eqnarray}
  \rh(t + \Delta t) - \rh(t) = -i \int\limits_t^{t + \Delta t} dt_1 \Tr_{\rm L}[ \Heff(t_1), U(t_1,t)\rho(t)U^\dagger(t_1,t)].
  \label{mid-step}
  \end{eqnarray}
We note that, $\Heff(t_1) = \hdd^I(t_1) + \hsl^I (t_1)$ and the commutation involving $\hl(t)$ goes to zero after taking the partial trace. 
Using the solution of the Schrodinger equation, we can write,
\begin{eqnarray}
\label{eq:2}
U(t_1) &=& \mathds{1} - i\int\limits_t^{t_1} dt_2 (\Heff(t_2) + \hl(t_2)) U(t_2)   
\end{eqnarray}
To construct the finite propagator, the following condition must be satisfied \cite{chakrabarti2018b}.
 \begin{itemize}
 \item The propagator contains only the first-order contribution of the perturbed Hamiltonian, $\Heff(t)$.
 \item It retains all order of the fluctuation Hamiltonian $\mathcal{H}_L(t)$, but upto a suitable time-interval.
 \end{itemize}
Here we note that, $[\Heff, \hl(t)] \neq 0$. We follow a particular truncation scheme that relies on the Neumann series solution of the Schrodinger equation. i.e.,
\begin{eqnarray}
U(t_1,t) &=& \mathds{1} -i \int\limits_t^{t_1} dt_2 H(t_2)  \biggl( \mathds{1} -i\int\limits_t^{t_2} dt_3 H(t_3)  \biggl( \mathds{1} -i \int\limits_t^{t_3} dt_4 H(t_4) \biggl( ....\nn
\end{eqnarray}
Here,  $ t_1>t_2>t_3>t_4.....$. Each integral in this series is smaller than the preceding one by $\vert H \vert \Delta t
<1$. A full solution involves painstakingly keeping track of all orders of $\heff(t)$ and
$\hl(t)$, which is indeed shown in the seminal work by Feynmann \cite{Feynman1951}.
However, we truncate using an ansatz that shows in Eq. (\ref{eq:2}), $U$ could be replaced
by $U_L$ $(U_L(t_2) = Te^{-i\int_t^{t_2} dt_3 \hl(t_3) dt_3})$ in the \emph{r.h.s} on the
ground that at  $t_2$ instances $\Heff$ has already been calculated and inclusion of
$\Heff$ from the expansion of $U$ will introduce a term having third order $\vert \hl
\times (\Heff)^2\vert \Delta t^3$, whose effect is negligible. It is rather surprising
that this truncation scheme provides a much closer result to experiments
\cite{chatterjee_nonlinearity_2020, saha_cascaded_2023, saha2024, chakrabarti2018a}.
We note that, the truncation scheme is strictly applicable for $\hsl
\neq 0$, otherwise the dynamics can be trivially decomposed in two different Hilbert 
spaces (i.e., system
and environment), and in such cases, no local interaction-induced dissipation can be
observed. However, using the form of a full propagator in the analysis and finding its
effect on the dynamics is a different open problem.
 
 The final form of the propagator is given by \cite{chakrabarti2018b}, 
\begin{eqnarray}
U(t_1) &=& \mathds{1} - i\int\limits_t^{t_1} dt_2 \Heff(t_2) U_L(t_2)   - i\int\limits_t^{t_1} dt_2 \hl(t_2) U_L(t_2)\nn\\
&=& U_L(t_1) - i\int\limits_t^{t_1} dt_2 \Heff(t_2) U_L(t_2) 
\label{propagator}
\end{eqnarray}
Using the form of the propagator (Eq. (\ref{propagator})) in Eq. \ref{mid-step}, we get, 
\begin{eqnarray}
\rh(t + \Delta t) - \rh(t) &=& -i \int\limits_t^{t+ \Delta t} dt_1 \Tr_{\rm L}[\Heff(t_1), U_L(t_1,t)\rho(t)U_L^\dagger(t_1,t)]\nn\\
&&  -\int\limits_t^{t+ \Delta t} dt_1 \int\limits_t^{t_1} dt_2 [\Heff(t_1), \Heff(t_2) U_L(t_2,t)\rho(t)U_L^\dagger(t_1,t)\nn\\
&& - U_L(t_2,t)\rho(t)U_L^\dagger(t_1,t) \Heff(t_2)]
\label{mid-term2}
\end{eqnarray}
Next, an ensemble average over the fluctuations is taken on both sides of the Eq. (\ref{mid-term2}). Using the Cumulant expansion as given by Kubo \cite{kubo_stochastic_1963}, we get,
\begin{eqnarray}
\overline{ U_L(t_1,t)\rho(t)U_L^\dagger(t_2,t)} &=& \rh(t) \otimes \rl \exp \left(- \frac{\vert t_1 - t_2 \vert}{\tau_c} \right)
\end{eqnarray}
We note that $t$ is the initial time of the coarse-grained interval $[t, t+\Delta t]$ and the system and bath are uncorrelated only at the initial time $t$ (i.e., Born approximation) \cite{cohen2004}. Putting the above formula in Eq. (\ref{mid-term2}) and using the coarse-grained  approximation as prescribed by Cohen-Tannoudji \etal \cite{cohen2004} $[\frac{d \rh}{dt} = \frac{\overline{\rh(t + \Delta t) - \rh(t)} }{\Delta t} ]$,   and further using the limit $\frac{\Delta t}{\tau_c} \to \infty$ (secular approximation), we get the final the final form of the FRQME, which is given by \cite{chakrabarti2018b},
 
\begin{eqnarray} \label{frqme1q}
\frac{d\rh}{dt}&=& -i \Tr_{\rm L}\Big[\heff(t),\rh \otimes\rl\Big]^{\rm sec}\nn\\
&&-\int\limits^{\infty}_0 d\tau \Tr_{\rm L}\Big[\heff(t),\Big[\heff(t-\tau),\rh \otimes\rl\Big]\Big]^{\rm
sec}e^{-\frac{\tau }{\tau_c}},
\end{eqnarray}
Here, the superscript `sec' denotes the secular
approximation. The presence of an exponential kernel $\left(\exp(-t/\tau_c)\right)$ results in a finite second-order contribution of $\hdd$ along with $\hsl$ and therefore goes beyond the independent rate approximation. As $\heff$ consists of $\hsl$ and $\hdd$, such a dynamical prescription successfully explains both the spin-spin $(\hdd-\hdd)$ and spin-lattice $(\hsl-\hsl)$ relaxation process using a single equation. We note that,  $\heff$ acts at two different time-instances `$t$' and `$t - \tau$'. In this interval, $\tau = \vert t - (t - \tau) \vert$, the bath dephases due to the fluctuations. The total evolution (full Hilbert space of the system and the environment) captures this dephasing, which results in the second-order dissipative effects of $\heff$. The above dynamical equation (Eq. (\ref{frqme1q})) can be reduced to the Gorini-Kossakowski-Sudarsan-Lindblad (GKSL) form, so here CPTP (complete positivity and trace preservation) holds \cite{chakrabarti2018b}.

\section*{References}
\bibliographystyle{unsrt}
\bibliography{reference1}

\end{document}